\documentclass[manuscript]{aastex}

\usepackage{amsmath, amssymb}

\shorttitle{IC X-RAY EMISSION LS 5039}
\shortauthors{Yamaguchi \& Takahara 2011}

\begin{document}


\title{INVERSE COMPTON SCATTERING MODEL FOR X-RAY EMISSION
OF THE GAMMA-RAY BINARY LS 5039}


\author{M. S. Yamaguchi and F. Takahara}
\affil{Department of Earth and Space Science, Graduate School of Science,
Osaka University, Toyonaka, Osaka, 560-0043, Japan}




\begin{abstract}
We propose a model for the gamma-ray binary LS 5039 in
which the X-ray emission is due to the inverse Compton
(IC) process instead of the synchrotron radiation.
Although the synchrotron model has been discussed in previous studies,
it requires a strong magnetic field which leads to a severe suppression
of the TeV gamma-ray flux in conflict with H.E.S.S. observations.
In this paper, we calculate the IC emission by low energy
electrons ($\gamma_e \lesssim 10^3$) in the Thomson regime.
We find that IC emission of the low energy electrons can explain the X-ray flux
and spectrum observed with \textit{Suzaku}
if the minimum Lorentz factor of injected electrons
$\gamma_{\text{min}}$ is around $10^3$.
In addition, we show that the \textit{Suzaku} light curve is well reproduced if
$\gamma_{\text{min}}$ varies in proportion to the \textit{Fermi} flux
when the distribution function of injected electrons at
higher energies is fixed.
We conclude that the emission from LS 5039 is well explained
by the model with the IC emission from electrons whose injection
properties are dependent on the orbital phase.
Since the X-ray flux is primarily determined by the total number of cooling
electrons, this conclusion is rather robust, although some mismatches
between the model and observations at the GeV band remain in the present
formulation.
\end{abstract}





\section{Introduction}
\label{intro}
Five gamma-ray binaries have been identified so far.
They are LS 5039 \citep{aha05a}, LSI +61$^{\circ}$ 303 \citep{alb06}, PSR B1259-63
\citep{aha05b},
HESS J0632+057 \citep{aha07,bon11}, and 1FGL J1018.6-5856 \citep{fer12}.
These systems consist of a compact object, a neutron star or a black hole, and
an OB star (LS 5039 and 1FGL J1018.6-5856 have an O star while the others have
a Be star).
The nature of the compact objects is still open except for PSR B1259-63
whose primary star is known to be a young pulsar. 

LS 5039 consists of a compact object and a massive star whose spectral
type is O6.5 V and whose mass is $M_* = 22.9^{+3.4}_{-2.9}\ M_{\odot}$,
and its binary period is 3.906 days \citep{cas05}. In addition,
the system has the eccentricity, $e = 0.24\pm0.08$ \citep{sar11},
and its separation changes from $\sim 2.6\ R_*$ at periastron
to $ \sim 4.2\ R_*$ at apastron, where $R_* = 9.3^{+0.7}_{-0.6}\ R_{\odot}$
represents the radius of the companion O star \citep{cas05}.
Spectra and light curves of LS 5039 have been reported in the energy band
of TeV gamma ray, GeV gamma ray, and X-ray, with H.E.S.S. array of
atmospheric Cherenkov telescopes \citep{aha06},
\textit{Fermi Gamma-ray Space Telescope} \citep{abd09}, and \textit{Suzaku}
\citep{tak09}, respectively. These observations show that the fluxes clearly
modulate with its binary period, which implies that the high energy emissions
are generated within the system.
In addition, we can see from the observational data that the X-ray flux
anti-correlates with the GeV gamma-ray flux and correlates with the TeV flux.
We note further that the photon index in the X-ray band is $\sim 1.5$,
which means that spectra of the electrons emitting X-rays have
a power-law index $\sim 2.0$.

Many authors have discussed the X-ray emission mechanism in LS 5039 assuming
that it is due to the synchrotron radiation
\citep{der06,dub06,par06,zha09,tak09,yam10,cer10}.
They modeled the system in a microquasar scenario \citep{der06,par06,zha09,tak09},
which includes jets from the accreting compact object, or in a colliding wind
scenario \citep{dub06}, which includes a relativistic pulsar wind
colliding with a stellar wind.
In each scenario, they obtained results more or less consistent with
the observations for all energy bands.

However, in explaining the X-ray emission as the synchrotron radiation,
there is a crucial problem.
The magnetic field of around 3 Gauss is required to reproduce the \textit{Suzaku}
spectrum \citep{tak09,yam10,cer10}, if the magnetic field is uniform and if
a relativistic motion of a radiating medium is neglected.
Under this magnetic field the cooling rate of the synchrotron process
dominates that of the inverse Compton (IC) process for electrons with
$\gtrsim$ 1 TeV.
This leads to a severe suppression of the flux in TeV range, so that
H.E.S.S. spectra cannot be reproduced, as shown in \citet{cer10}.

On the other hand, the work trying to explain X-ray emission as
IC process has been limited.
\citet{bos05} suggested that the X-ray emission was due to the IC (and/or
synchrotron) process from the jets and that the orbital variation of
the accretion rate was responsible for the X-ray modulation before H.E.S.S. and
\textit{Suzaku} observations were reported.
After this study, the IC model for the X-ray emission has not been discussed.
In the light of the new observational data,
we revisit the IC model to reproduce the spectra and the orbital
modulation of the X-ray emission.
In this paper we aim to find the condition to reproduce
the \textit{Suzaku} data when we take the same parameters as
in our previous paper, \citet{yam10} (hereafter YT10).

We describe the model in Section \ref{lowe}, and show obtained results in
Section \ref{leres}, where we demonstrate that the IC model can well reproduce
the X-ray observations if certain conditions are satisfied.
We discuss remaining problems and complexities
in Section \ref{dis}.
Finally, we summarize the study in Section \ref{sum}.

\section{IC emission model for X-ray emission}
\label{lowe}
\subsection{Model overview}
\label{ov}
We assume that high energy electrons are accelerated with an isotropic
distribution at the position of the compact object.
Here, we do not specify the acceleration mechanism, i.e.,
the microquasar model or the wind collision model.
Since the electrons are in the stellar radiation field, they lose energy by
IC process. We neglect the synchrotron cooling by assuming that the magnetic
energy density is much lower than the radiation energy density.
This cooling process leads to modification of the electron distribution as
given in the next section. 
On the other hand, the electrons emit X-rays and gamma rays by scattering off the
stellar radiation.
Since the stellar radiation field is anisotropic, the high energy emission is
also anisotropic resulting from anisotropic IC scattering.
This anisotropy affects the modulation of the emission.

Here, we focus on low energy electrons whose IC photons are not subject to
$\gamma \gamma$ absorption, which have been neglected in previous works.
In YT10, we injected electrons
with $3\times 10^3 < \gamma < 10^8$ and calculated IC scattering,
the $\gamma \gamma$ absorption and subsequent cascade process.
In addition, we calculated the IC cooling down to $\gamma_e = 3\times 10^3$.
Thus, we focused on the emission from higher energy electrons.
\citet{tak09} also concentrated on high energy IC emission ($\gtrsim$ 1 GeV).
In this paper we calculate the emission from electrons with lower energy,
which emit X-rays and MeV gamma rays through the IC process.

\subsection{Electron distribution}
\label{eledi}
The energy distribution of injected electrons $n_{\text{inj}}(\gamma)$
is assumed as
\begin{equation}
n_{\text{inj}}(\gamma) \propto \gamma^{-p} \ \ \ \text{for }
\gamma_{\text{min}} < \gamma < \gamma_{\text{max}},
\label{injele}
\end{equation}
where $\gamma_{\text{min}} \text{ and }\gamma_{\text{max}}$ are the minimum
and maximum Lorentz factors of the injected electrons, respectively.
Because gamma rays up to 30 TeV have been detected from LS 5039, we assume
$\gamma_{\text{max}} = 10^8$.
However, now we are interested in X-ray emission.
Moreover, we have calculated GeV to TeV emission due to IC process by the
Monte Carlo method and
successfully reproduced the observations in YT10.
Therefore, in this paper we calculate the IC emission from the electrons
with $\gamma < \gamma_1 \equiv 3\times 10^3$, which corresponds to
the minimum Lorentz factor in YT10.
In addition, we adopt the same power-law index of the injected electrons as in YT10,
i.e., $p = 2.5$, and we assume that $p$ and the coefficient in Equation
(\ref{injele}) are constant in the whole orbit.

The electrons severely lose energy due to the strong radiation field by the star.
Given the energy density of the stellar radiation, the cooling time by IC process
in the Thomson regime is
\begin{equation}
t_{\text{IC,T}} \sim 3\times 10^4 \gamma^{-1} \left( \frac{a}{a_{\text{peri}}}
\right)^2 \text{s,}
\label{tict}
\end{equation}
where $a$ represents the binary separation and 'peri' means periastron and $\gamma$
is the Lorentz factor of electrons.
Equation (\ref{tict}) implies that injected electrons completely cool near
the periastron in $T_{\text{orb}}/10 \sim 3\times 10^4$ s, where $T_{\text{orb}}$
is the orbital period,
while at the apastron electrons cool down to $\gamma_\text{c} \sim 4$
in $T_{\text{orb}}/10$, where $\gamma_\text{c}$ means the Lorentz factor
of cooled electrons.

Therefore, the electron distribution is modified by IC cooling. As a result,
if $\gamma_{\text{min}} > \gamma_\text{c}$
the distribution in the steady state is given by the following form,
\begin{equation}
n(\gamma) \propto \begin{cases}
 \ \  \gamma^{-2} & \text{for } \gamma_{\text{c}} < \gamma < \gamma_{\text{min}} \\
 \ \  \gamma^{-(p+1)} & \text{for } \gamma_{\text{min}} < \gamma < \gamma_1
		  \end{cases}.
\label{ssele}
\end{equation}
The normalization of this distribution is determined by equating the number
of electrons with $\gamma_1$ to that obtained at $\gamma_1$ by the Monte Carlo
calculation in YT10. Therefore the electron distribution of Equation
(\ref{ssele}) is continuously joined to that of YT10. 
We note that $n(\gamma)$ in YT10 is just a boundary condition in this paper,
so that the calculation below does not affect so much the GeV spectra.
At the same time some mismatches between the model and observed spectra
at the GeV band remain.

\subsection{Parameter $\gamma_{\text{min}}$}
\label{gmin}
We assume that $\gamma_{\text{min}}$ is a free parameter
so that we fit the IC emission spectra from these electrons
with the \textit{Suzaku} data.
The X-ray flux observed by \textit{Suzaku} anti-correlates with
the \textit{Fermi} flux.
If we assume that $\gamma_{\text{min}}$ is constant
with the orbital phase, we expect that the fluxes at the two energy bands
correlate with each other. This is because the orbital modulation is wholly
due to anisotropic IC.
In this case, the light curve of \textit{Fermi} is reproduced by the anisotropy
of IC process (see YT10), but that of \textit{Suzaku} cannot be reproduced.
Therefore, we should assume that $\gamma_{\text{min}}$ varies
with the orbital phase so that the X-ray light curve fits the observed one. 
We note that this assumption is completely ad hoc at this stage, but it is possible
that $\gamma_{\text{min}}$ varies with the orbital phase because the condition
of the electron injection could change with the orbital separation.

Here, we discuss the relation between the X-ray flux and $\gamma_{\text{min}}$
in order to understand how we should set the variation of $\gamma_{\text{min}}$.
We define $F_{\text{x,}0}$ and $n_0(\gamma_{\text{x}})$ as the IC X-ray flux and
the electron number spectrum at $\gamma_{\text{x}}$, respectively,
expected when
$\gamma_{\text{min}}$ is constant with the orbital phase,
where $\gamma_{\text{x}}$ represents the Lorentz factor of electrons emitting X-ray,
i.e., $\gamma_{\text{x}} \sim 10$.
We note that $F_{\text{x,}0}$ shows the orbital modulation similar to 
the \textit{Fermi} flux $F_{\text{GeV}}$ because the modulation of $F_{\text{x,}0}$
correlates with that of the IC flux in the GeV band and because this GeV flux well reproduces
the \textit{Fermi} flux (see YT10).
When $\gamma_{\text{min}}$ varies with the orbital phase, the expected X-ray flux
$F_{\text{x}}$ and the electron number spectrum $n(\gamma_{\text{x}})$ satisfy the relation
\begin{equation}
F_{\text{x}} = F_{\text{x,}0} \ \frac{n(\gamma_{\text{x}})}{n_0(\gamma_{\text{x}})}.
\label{fx}
\end{equation}
Using Equation (\ref{ssele}), we obtain
\begin{equation}
n(\gamma_{\text{x}})
= n(\gamma_1) \left( \frac{\gamma_{\text{min}}} {\gamma_1} \right)^{-(p+1)}
\left( \frac{\gamma_{\text{x}}} {\gamma_{\text{min}}} \right)^{-2}.
\label{nx}
\end{equation}
Since we assume that the electron number at $\gamma_1$ varies in the same way as
that in YT10, we set
\begin{equation}
n(\gamma_1) = n_0(\gamma_1).
\end{equation}
When $\gamma_{\text{min}}$ is constant during the whole orbit, $n_0(\gamma_1)$
shows the same modulation as $n_0(\gamma_{\text{x}})$, so that we obtain
\begin{equation}
n(\gamma_1) \propto n_0(\gamma_{\text{x}}).
\end{equation}
Therefore, Equation (\ref{nx}) leads to
\begin{equation}
n(\gamma_{\text{x}}) \propto n_0(\gamma_{\text{x}}) \gamma_{\text{min}}^{1-p}
\end{equation}
As a result, Equation (\ref{fx}) reduces to
\begin{equation}
F_{\text{x}} \propto F_{\text{x,}0} \ \gamma_{\text{min}}^{-1.5},
\label{fxg}
\end{equation}
where we set $p=2.5$.
Therefore, we should increase $\gamma_{\text{min}}$ when $F_{\text{x,}0}
(F_{\text{GeV}})$ increases and vice versa, in order to reproduce
the observed orbital modulation at X-rays.

\subsection{Assumptions for the calculation}
\label{cal}
We assume that the electrons which follow the energy distribution given by Equation
(\ref{ssele}) are in a steady state for a given orbital phase
because $\gamma_{\text{x}} > \gamma_{\text{c}}$.
The stellar radiation is assumed to take a black-body distribution.
The typical energy is 10 eV, so that IC scattering occurs
in the Thomson regime for electrons $\gamma < \gamma_1$.
Therefore, we calculate numerically the IC spectra in the Thomson regime,
which is done in the same manner as in \citet{dub08}.
Here, we neglect the effect of secondary electrons created by the cascade process.

\section{Calculated Results}
\label{leres}
\begin{figure}
\begin{center}
\includegraphics[height=13cm,angle=270,clip]{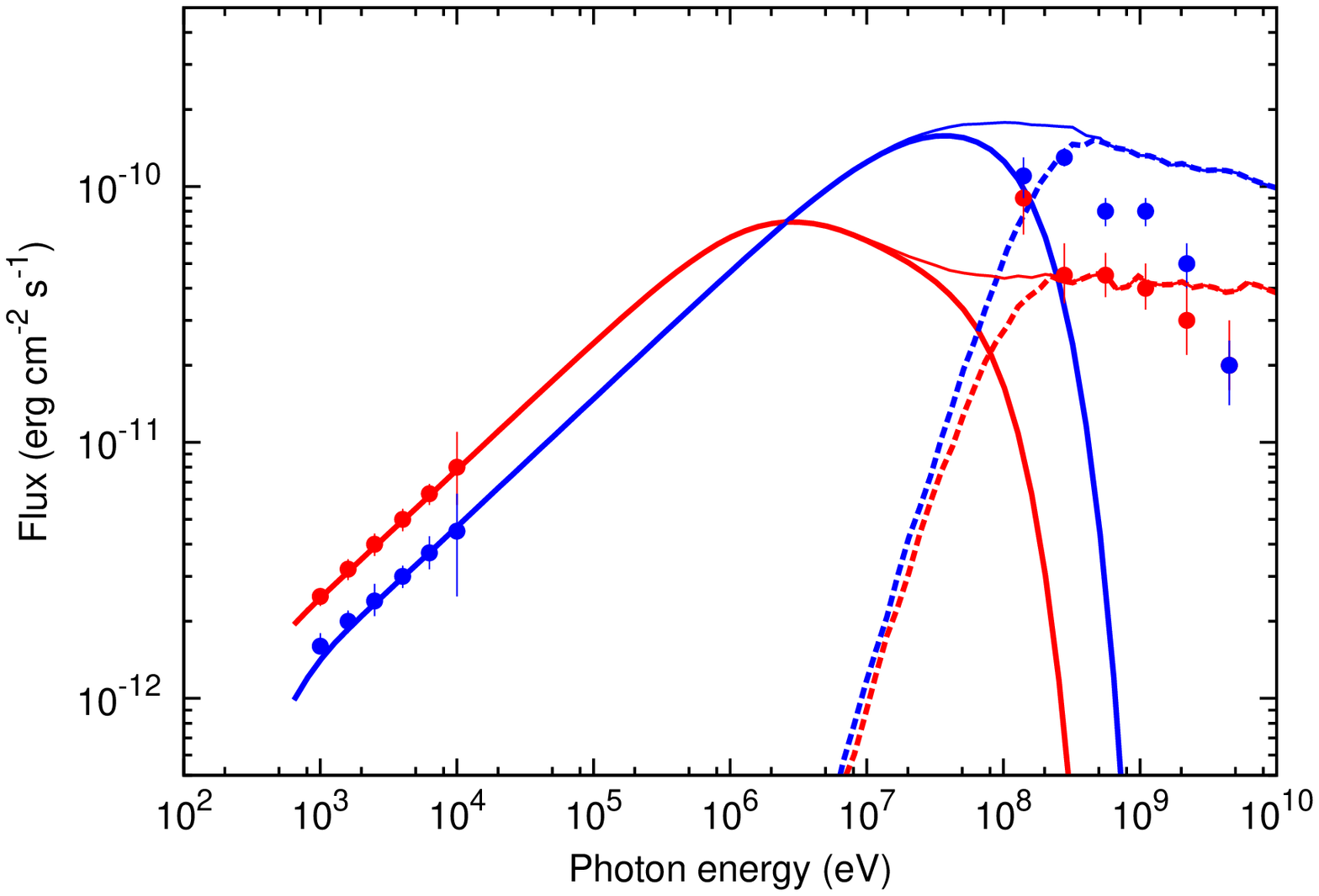}
\caption{IC spectra by low energy electrons ($\gamma < 3\times 10^3$).
The spectra are averaged in the two orbital phases, around inferior conjunction 0.45-0.9
(red lines and circles) and around superior conjunction 0.9-0.45 (blue lines and circles).
 The observed spectra are shown by circles with error bar for
\textit{Suzaku} \citep[XIS,][]{tak09} and \textit{Fermi} \citep{abd09}.
IC spectra obtained by Monte Carlo method for $\gamma > 3\times 10^3$ are also shown
with dashed curves, which are equivalent to those in Figure 6 of YT10.
We change $\gamma_{\text{min}}$ as a function of the orbital phase (Figure \ref{gevfit})
so that the calculated
X-ray light curve fits the observed one as shown in Figure \ref{lcfit}.
Thin solid lines show the sums of the spectra from low and high energy electrons in each phase.}
\label{x-ic}
\end{center}
\end{figure}

\begin{figure}
\begin{center}
\includegraphics[height=13cm,angle=270,clip]{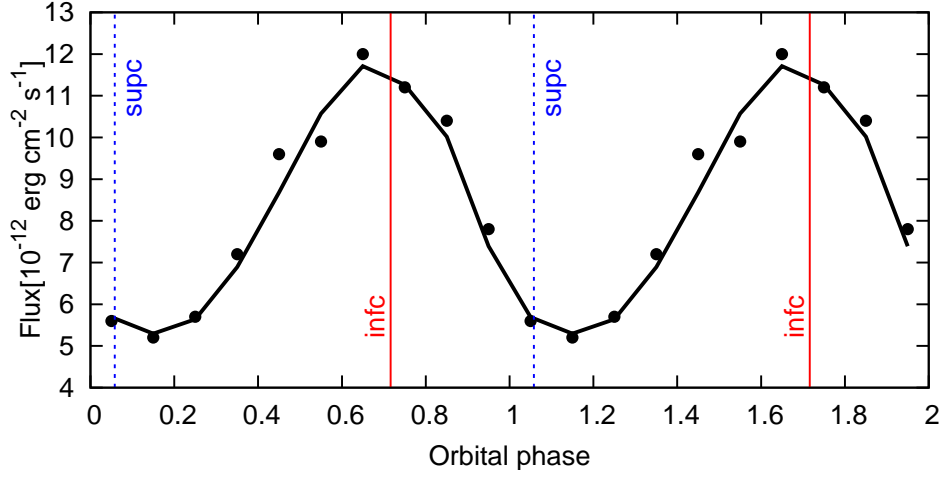}
\caption{Calculated (solid line) and observed (circles) X-ray light curves for the case
as in Figure \ref{x-ic}.
The vertical lines show the superior conjunction (supc) and inferior conjunction (infc).}
\label{lcfit}
\end{center}
\end{figure}

In Figure \ref{x-ic}, the resultant spectra from low energy electrons (solid lines),
the IC spectra from high energy electrons obtained in YT10 (dashed lines) and
their sums (thin solid lines) are shown,
where the spectra are averaged in the two orbital phases,
around the inferior conjunction $\phi =$0.45-0.9 (red lines and circles) and
around the superior conjunction $\phi =$0.9-0.45 (blue lines and circles).
Here, $\phi$ means the orbital phase, and $\phi = 0$ represents the periastron.
The orbital variation of $\gamma_{\text{min}}$
is given so that the light curve of the observation with \textit{Suzaku} is reproduced
(Figure \ref{lcfit}).
We set the minimum Lorentz factor of cooling electrons as $\gamma_{\text{c}} = 5$,
so that the cooling time of electrons with this Lorentz factor is shorter than
a tenth of the orbital period at any orbital phase (see Section \ref{eledi}).
The inclination angle is taken as $i = 30^{\circ}$ so that this model is
consistent with YT10. The orbital parameters are taken from \citet{sar11}.
The results do not change so much even if we take other allowable parameter sets
\citep[e.g. those given in][]{ara09}.

\begin{figure}
\begin{center}
\includegraphics[height=13cm,angle=270,clip]{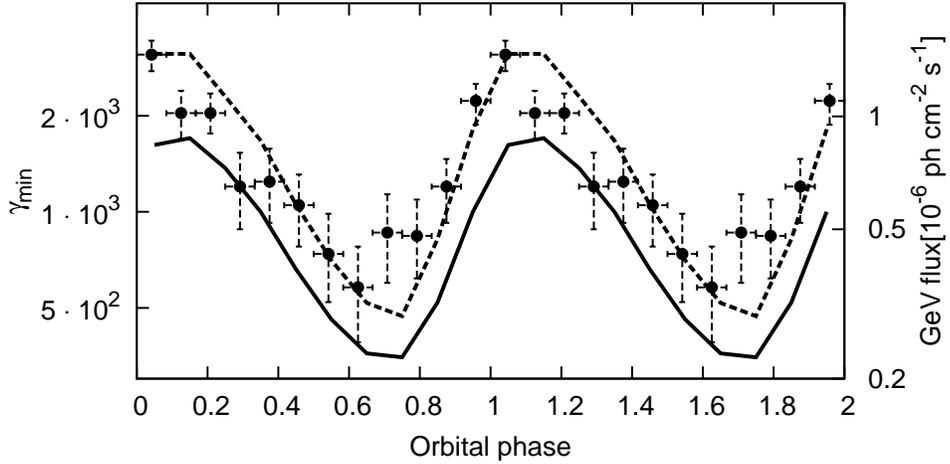}
\caption{Adopted phase dependence of $\gamma_{\text{min}}$ (solid line) for the case of
Figures \ref{x-ic} and \ref{lcfit}.
Dashed line shows $F_{\text{x,}0}$,
and circles with dashed error bar show the observed GeV light curve.}
\label{gevfit}
\end{center}
\end{figure}

We determine the orbital variation of $\gamma_{\text{min}}$ by fitting
the calculated light curve to the observed one as shown in Figure \ref{lcfit}.
The comparison of the variation of $\gamma_{\text{min}}$ with the \textit{Fermi}
light curve is shown in Figure \ref{gevfit},
where $\gamma_{\text{min}}$ is shown with a solid line, $F_{\text{x,}0}$ is shown
with a dashed line, and the flux obtained with \textit{Fermi} is shown by crosses
with dashed error bars.
Figure \ref{gevfit} indicates that $\gamma_{\text{min}}$ obtained
from the fitting to the X-ray band shows almost the same variation as the IC flux with
the constant injection, that is,
\begin{equation}
\gamma_{\text{min}} \propto F_{\text{x,}0} \sim F_{\text{GeV}}.
\label{gmic}
\end{equation}
In other words, if we assume the variability of $\gamma_{\text{min}}$ as
Equation (\ref{gmic}), we can reproduce the observation with \textit{Suzaku}.
The reason why the fitting to the data of \textit{Suzaku} yields such a simple
relation is that the observed fluxes of X-ray and GeV bands are related as
\begin{equation}
F_{\text{x}} \propto F_{\text{GeV}}^{-\frac{1}{2}}.
\label{sufe}
\end{equation}
In addition to the orbital variation of $\gamma_{\text{min}}$, 
we note that this result gives a lower limit of $\gamma_{\text{min}}$.
If $\gamma_{\text{min}}$ is lower than $\sim 10^3$, calculated flux overpredicts
the \textit{Suzaku} one.

\section{Discussion}
\label{dis}
We have shown in the previous section that the orbital modulation of
the X-ray emission from LS 5039 is reproduced by the IC process if we give
the ad hoc relation of Equation (\ref{gmic}).
However, it is not clear if such a dependence is expected.
It is hard to assume that the minimum Lorentz factor
of injected electrons varies in tandem with the line of sight. Nevertheless,
since the existence of cooled electrons is inevitable, it is natural
that the low energy electrons with $\gamma \sim 10$ is responsible for
the X-ray emission from LS 5039.

In the model of this paper, the X-ray modulation is mainly due to
the variation of the number of cooled electrons.
Since the number of cooled electrons is determined by the injection rate
$\int n_{\text{inj}}(\gamma)d\gamma$,
the X-ray modulation is controlled by the injection rate.
In this paper, we assume that $p$ and the coefficient in Equation
(\ref{injele}) are constant in the whole orbit
while $\gamma_{\text{min}}$ modulates.
Here we note that the injection rate changes with the orbital motion.
That is, the X-ray modulation is adjusted by making $\gamma_{\text{min}}$
change in this paper.
Alternatively, we may adjust the X-ray modulation by changing $p$ and the coefficient
because these two parameters are also related to the injection rate.

While the \textit{Suzaku} spectra are well fitted in Figure \ref{x-ic},
the calculated spectra in the GeV band match worse the observation
since we adopt the YT10 data as higher energy spectra.
The \textit{Fermi} spectrum around the superior conjunction is now somewhat
overpredicted even for the lowest two bins of \textit{Fermi} spectrum.
This means that the amplitude of the modulation in the GeV band is
slightly overpredicted when the contribution from low energy electrons is considered.
Therefore, in order to reproduce the \textit{Fermi} observation we should
set the smaller amplitude of the calculated light curve than the current model,
which corresponds to a smaller inclination angle.
Smaller inclination angle means a larger mass of the compact object,
strengthening the suggestion in YT10 that it is a black hole.
This modification of the inclination angle does not affect so much
Equation (\ref{gmic}) although the amplitude of $F_{\text{x,}0}$ and
thereby that of $\gamma_{\text{min}}$ are slightly modified.
We note that in this model TeV gamma-ray can be emitted by the IC process
since this model is compatible with the model in YT10.

Here, we mention the effect of the escape of electrons on the model.
We assume in this paper that electrons cool down to $\gamma_{\text{c}}$
without escaping.
However, when taking into account the hydrodynamics of the system, e.g.,
the outflow of the jet in the microquasar scenario, if we assume that
electrons are advected by the flow, they cannot cool radiatively so efficiently.
For example, if electrons are advected at the light speed,
the crossing time in the system $t_{\text{cross}}$ is $\sim 100$ s.
Since the electrons cool during this crossing time, $\gamma_{\text{c}}$ is
determined by
\begin{equation}
t_{\text{IC,T}} = t_{\text{cross}},
\label{gesc}
\end{equation}
where $t_{\text{IC,T}}$ is given by Equation (\ref{tict}).
Thus, we obtain $\gamma_{\text{c}} \sim 300$, which means that
electrons do not emit X-rays through the IC scattering if advection is very fast.
This issue may be solved in two ways.
One is the adiabatic cooling where the spatial extent of electron
confinement increases more than two orders of magnitude.
The other is that electron injection itself continues to as low as
$\gamma_{\text{min}} \sim 10$ with an index 2.0 below a break,
then the photon index in the X-ray band may be reproduced.
Since $\gamma_{\text{c}} > \gamma_{\text{min}}$, electrons emitting X-rays
are not affected by the IC cooling.
In these scenarios, the X-ray flux may modulate due to not only the anisotropic
IC process but also the orbital variation of $\gamma_{\text{c}}$,
but further study is required to reproduce the \textit{Suzaku} observations.

We comment shortly models based on the synchrotron emission.
If we ascribe the X-ray emission to the synchrotron radiation,
the TeV observation cannot be reproduced as shown in \citet{cer10}.
This discrepancy may be solved by a model in which the magnetic field
is non-uniform.
For example, X-rays are emitted in a compact region where the magnetic field is
more than 3 Gauss and the electrons are populated in energies lower than 1 TeV,
while TeV IC emission by the higher energy electrons takes place
in an extended region with magnetic field strength as small as 0.1 Gauss.
As an alternative scenario addressing this problem, we could consider
a situation where the radiative
matter moves along near the line of sight at relativistic velocity. 
In this scenario, the \textit{Suzaku} spectra would be reproduced under the
magnetic field less than 1 Gauss because of the relativistic beaming,
and a severe suppression in the TeV band could be avoidable.

\citet{tak09} proposed the model in which the X-ray emission is explained
by the synchrotron radiation.
They assume that the adiabatic cooling is more efficient than the radiative
cooling, so that the X-ray flux is proportional
to the adiabatic cooling time.
Thus, the X-ray light curve can be reproduced if one give properly the variation of
the adiabatic cooling time.
Moreover, since the adiabatic cooling does not change the power-law index
of the electron distribution, the spectral index in the X-ray band
is also reproduced if one give $n_{\text{inj}}(\gamma) \propto \gamma^{-2}$.
In addition, they show that calculated TeV spectra match the H.E.S.S. observation.
The suppression in the TeV band is avoided in this model
because high energy electrons cool not radiatively but adiabatically, 
so that they are not affected by the synchrotron cooling.

Many authors have constructed models in which the parameters of injected
electrons vary with the orbital motion to reproduce the observations.
\citet{par06} and \citet{zha09} calculated the emission from the jet
in the microquasar scenario and showed broad band spectra from radio to TeV
gamma-ray band.
\citet{par06} assumed that the injection rate of the electron
distribution varied with the orbital motion due to the change in the accretion
rate, and investigated the orbital modulation in each energy band.
However, their calculation was based on the data before \textit{Suzaku}.
\citet{zha09} assumed that the electron power-law index and the accretion rate,
which is equivalent to the injection rate, showed two values for two orbital
phase, superior and inferior conjunctions, respectively.
As a result, the \textit{Suzaku} spectra are well reproduced but
they did not mention the orbital modulation for any energy band.
\citet{sie08a} and \citet{sie08b} calculated the IC scattering from electrons
in the pulsar wind in the wind collision scenario and showed that their
spectra and light curves reproduced the H.E.S.S. data.
They assumed that the distribution
of the injected electrons showed two different power-law indices for the
two phase intervals, respectively, in order to reproduce the H.E.S.S. spectra.
However, they have not investigated X-ray emission.

\citet{bed11} assumed that in the wind collision scenario electrons
were accelerated in two shocks separated by a contact discontinuity.
He have shown that the IC emission from the pulsar-side shock and
the star-side shock accounts for TeV emission and GeV emission, respectively.
In his model the maximum energy of electrons varied with
the orbital motion, and the injection rate was constant.
As a result, the normalization of the electron distribution is changed,
so that X-ray emission by the synchrotron radiation modulates,
although the flux varies inversely to the \textit{Suzaku} data.

The crucial difference between the IC model and the synchrotron model is 
in the MeV spectra.
The IC model in this paper predicts that there is no dip in MeV spectra.
In contrast, the synchrotron model predicts that there is a dip in MeV spectra
(\citet{tak09}, YT10, \citet{cer10}).
Therefore, future observations in the MeV band will judge which model is realized.

\section{Summary}
\label{sum}
We assume that X-ray emission from LS 5039 is due to the IC scattering of electrons
with $\gamma \sim 10$.
We calculate the anisotropic IC emission from electrons with the Lorentz factor
$5 < \gamma < 3\times 10^3$ in the Thomson regime.
As a result, we have found that
we can reproduce \textit{Suzaku} flux if the minimum Lorentz factor of injected electrons
$\gamma_{\text{min}} \sim 10^3$.
Moreover, the X-ray light curve is well reproduced if
a simple relation $\gamma_{\text{min}} \propto F_{\text{GeV}}$ is satisfied,
where $F_{\text{GeV}}$ represents the GeV flux.
We have obtained these results based on YT10,
where the index and the coefficient of the power-law distribution for
injected electrons do not change with respect to the orbital phase.
We may obtain different results when we adopt the model in which the index
and the coefficient vary, but \textit{Suzaku} data can be reproduced
by adjusting these parameters, because the X-ray flux variation is
primarily determined by the number of low energy electrons.
Thus, we suggest that the X-ray emission obtained with
\textit{Suzaku} can be explained by the IC emission.
However, we need more investigation for explaining the \textit{Fermi} data.
This model will be justified or rejected by future observations in the MeV range.


\acknowledgments
M.S.Y. 
is supported by a Grant-in-Aid for JSPS Research Fellowships for Young Scientists
(A231860).

\clearpage



%

\clearpage

\end{document}